\documentclass[journal]{IEEEtran}
\usepackage{cite}

\usepackage{graphicx}
\ifCLASSINFOpdf
\else
\fi
\usepackage{amsmath}
\usepackage{amsfonts}
\usepackage{amssymb}
\begin{document}
%
\title{Convolutional Dictionary Learning Based Hybrid-Field Channel Estimation for XL-RIS-Aided Massive MIMO Systems}
%
%
%
\author{Peicong~Zheng, Xuantao~Lyu, Ye~Wang,~\IEEEmembership{Member,~IEEE}, and Yi~Gong,~\IEEEmembership{Senior Member,~IEEE}
}

\maketitle 

\begin{abstract}
Extremely large reconfigurable intelligent surface (XL-RIS) is emerging as a promising key technology for 6G systems. To exploit XL-RIS's full potential, accurate channel estimation is essential.
This paper investigates channel estimation in XL-RIS-aided massive MIMO systems under hybrid-field scenarios where far-field and near-field channels coexist.
We formulate this problem using dictionary learning, which allows for joint optimization of the dictionary and estimated channel. 
To handle the high-dimensional nature of XL-RIS channels, we specifically adopt a convolutional dictionary learning (CDL) formulation.
The CDL formulation is cast as a bilevel optimization problem, which we solve using a gradient-based approach.
To address the challenge of computing the gradient of the upper-level objective, we introduce an unrolled optimization method based on proximal gradient descent (PGD) and its special case, the iterative soft-thresholding algorithm (ISTA). 
We propose two neural network architectures, Convolutional ISTA-Net and its enhanced version Convolutional ISTA-Net+, for end-to-end optimization of the CDL.
To overcome the limitations of linear convolutional filters in capturing complex hybrid-field channel structures, we propose the CNN-CDL approach, which enhances PGD by replacing linear convolution filters with CNN blocks in its gradient descent step, employing a learnable proximal mapping module in its proximal mapping step, and incorporating cross-layer feature integration.
Simulation results demonstrate the effectiveness of the proposed methods for channel estimation in hybrid-field XL-RIS systems.
\end{abstract}


%
\IEEEpeerreviewmaketitle

\section{Introduction}
%
%
%
%
Reconfigurable intelligent surfaces (RIS) have emerged as a promising technology to enhance the performance and coverage of wireless communication systems\cite{9424177}. 
By leveraging a large number of low-cost passive reflecting elements, RIS can dynamically alter the phase of incident electromagnetic waves, thereby creating favorable channel conditions for communication. 
To fully leverage the potential of RIS in enhancing wireless communication performance, accurate channel state information (CSI) is crucial\cite{9722893}.
However, acquiring accurate CSI in RIS-aided systems presents significant practical challenges, primarily due to two main reasons\cite{9771077}.
Firstly, the passive nature of RIS elements limits their ability to participate in baseband signal processing.
Secondly, the large number of RIS elements significantly increases the pilot overhead required for channel estimation.

\subsection{Prior Work}
Channel estimation research for RIS-aided communication systems has primarily focused on the cascaded channel\cite{9732214}, which is the product of the BS-RIS channel and the RIS-user channel.
To address the challenge of high pilot overhead in estimating this cascaded channel, researchers proposed compressed sensing (CS)-based methods \cite{9328485,10053657,9881980,9103231,9475488,9521836}. 
These methods leverage the sparsity of the cascaded channel, resulting in reduced pilot requirements.
Specifically, leveraging the sparse nature of millimeter wave (mmWave) channels, the cascaded channel estimation is formulated as a sparse signal recovery problem\cite{spawc}.
Based on this formulation, various algorithms have been applied to solve the problem, including orthogonal matching pursuit (OMP) based algorithms \cite{9328485,10053657}, message passing based algorithms \cite{9881980,9103231}, and the least absolute shrinkage and selection operator based algorithms \cite{9475488,9521836}.

The evolution of RIS is moving towards extremely large-scale RIS (XL-RIS), which are expected to play a crucial role in future 6G communications\cite{9810144}.
Conventional RIS typically employ tens to a few hundred elements.
In contrast, XL-RIS systems are characterized by a substantial increase in the number of elements, potentially scaling from hundreds to thousands (e.g., from 256 to 1024 or more). 
This expansion in the number of elements leads to an enlarged array aperture, altering the electromagnetic field properties\cite{10496996}.
As the array size increases, the Rayleigh distance, which defines the boundary between near-field and far-field regions, extends proportionally.
Consequently, in XL-RIS systems, some scatterers previously located in the far-field of conventional RIS now fall within the near-field region.
This shift from far-field to near-field propagation necessitates a fundamental change in wavefront modeling.
In far-field propagation with conventional RIS, the wavefront can be simply modeled under the planar wavefront assumption, which only depends on the angle of departure/arrival (AoD/AoA) of the channel.
This allows the discrete Fourier transform (DFT) matrix to serve as the dictionary matrix, enabling a sparse representation of the channel in the angular domain\cite{10178011}.
However, in the near-field propagation of XL-RIS, the wavefront should be accurately modeled under the spherical wavefront assumption, which is determined not only by the AoD/AoA but also by the distance between the XL-RIS and the scatterers\cite{10379539}.
Therefore, the aforementioned channel estimation methods, which are designed for far-field scenarios in conventional RIS, may no longer be directly applicable for XL-RIS.

The channel estimation for XL-RIS builds upon research in extremely large multiple-input multiple-output (XL-MIMO) systems.
Due to the near-field spherical wavefront property, the channel sparsity in the angular domain is not achievable. 
To address this challenge, the authors in \cite{9693928} proposed a polar-domain dictionary, which simultaneously accounts for both the angular and distance information.
This approach enables the near-field channel to exhibit sparsity in the polar domain.
Building on these insights from XL-MIMO research, several studies have explored the application of polar-domain sparse representation to XL-RIS channel estimation.
In \cite{10149498}, the XL-RIS channel estimation problem was transformed into a sparse recovery problem by utilizing polar-domain sparsity. 
This work jointly estimated the channel of all subcarriers using the block orthogonal least squares algorithm, leveraging the common support property.
A decoupled polar-angular-domain strategy was introduced in \cite{10081022} for XL-RIS near-field channel estimation. This strategy combines a distributed CS framework for angular parameter recovery with a line-of-sight (LOS)-based polar-domain dictionary for polar parameter recovery.
A two-stage learning-based scheme for channel estimation in near-field XL-RIS systems was developed in \cite{10464973}, leveraging the cascaded channel's dual-structured sparsity.
In \cite{10153711}, a channel estimation method for XL-RIS systems using fast sparse Bayesian learning was proposed, focusing on the limited visibility regions between the RIS and individual users.
For XL-RIS-aided indoor systems with LOS-only paths between XL-RIS and users, the work in \cite{10077727} addressed channel estimation by proposing a U-shaped multilayer perceptron network to reconstruct non-stationary channels caused by limited visibility regions.

The aforementioned studies primarily focus on scenarios where the XL-RIS and users are in the near-field region.
However, in practical deployments, the propagation environment often leads to hybrid-field scenarios, where near-field and far-field propagation coexist between the XL-RIS and users.
The authors in \cite{9598863} proposed a hybrid-field channel estimation scheme for XL-MIMO systems.
Their approach separately estimates far-field and near-field path components using different dictionaries. 
Specifically, the far-field path components are estimated by leveraging their sparsity in the angle domain using a DFT dictionary, while the near-field path components are estimated using a polar domain dictionary to exploit their sparsity in the polar domain.
Based on this idea of separately estimating far-field and near-field path components, subsequent research has developed various improved methods, such as the support detection OMP \cite{9940281}, the convolutional autoencoder OMP \cite{10013010}, and the stochastic gradient pursuit algorithm\cite{10546479}.
Unlike methods that estimate near-field and far-field paths separately, a unified approach was proposed in \cite{10143629}  using a DFT dictionary to capture both near-field and far-field characteristics simultaneously and using deep learning networks for hybrid-field channel estimation in XL-MIMO systems.
These methods rely on predefined dictionaries, such as DFT and polar domain dictionaries. 
While these methods have demonstrated effectiveness, they suffer from modeling errors due to the inherent limitations of predefined dictionaries in accurately representing complex hybrid-field channel structures.
Furthermore, for XL-RIS systems, the design of an appropriate dictionary for the cascaded channel in hybrid-field scenarios poses significant challenges due to the intricate interplay between near-field and far-field propagation effects.

To overcome these limitations, an alternative approach is to learn a dictionary that can better capture the underlying structure of the cascaded channel in hybrid-field scenarios. 
This concept of dictionary learning (DL) for channel sparse representation and estimation was initially proposed in \cite{8383706} for massive MIMO systems. 
In this work, an overcomplete dictionary was learned from channel measurement data, offering a more adaptive alternative to predefined DFT dictionaries.
Building on the idea of DL, the authors in \cite{9186336,10054604} extended the application of dictionary learning to wideband mmWave MIMO systems with hardware impairments and beam squint.
In \cite{10197339}, dictionary learning methods were developed for channel estimation in massive MIMO system, exploiting both channel sparsity and angle reciprocity between uplink and downlink transmissions.
However, these existing DL methods are primarily designed for uniform linear arrays, whereas XL-RIS typically employs uniform planar arrays to accommodate more antennas within a limited aperture area. 
Moreover, the computational complexity of these methods becomes prohibitive for XL-RIS systems due to the need for an overcomplete dictionary, where the number of basis vectors in the dictionary is large and the dimension of each basis vector is proportional to the number of elements in the XL-RIS.
Additionally, the alternating optimization approach used in these methods, which separates dictionary learning and channel estimation, might not be optimal for the complex hybrid-field cascaded channels in XL-RIS systems.
\subsection{Our Contributions}
In light of the aforementioned challenges and limitations, the main contributions of this paper are as follows:
\begin{itemize}
    \item We investigate channel estimation in XL-RIS-aided massive MIMO systems, addressing the challenges posed by hybrid-field scenario.
    We formulate the channel estimation problem within the dictionary learning framework, which allows adaptive learning of the dictionary and estimation of the channel, overcoming the limitations of relying on fixed dictionaries. 
    \item To handle the high-dimensional nature of XL-RIS channels, we adopt a convolutional dictionary learning (CDL) approach.
    This approach exploits the properties of convolution to capture the structure of channel over its entire high-dimensional space, thereby reducing the computational burden in conventional dictionary learning approaches.
    \item The CDL is cast as a bilevel optimization problem. 
    We adopt a gradient-based approach to solve it.
    A key challenge in this approach is computing the gradient of the upper-level objective, as the solution of the lower-level optimization problem cannot be obtained explicitly.
    We overcome this challenge by introducing an unrolled optimization method that approximates the lower-level solution using an iterative optimization algorithm.
    Specifically, we leverage the proximal gradient descent (PGD) algorithm, which simplifies to the iterative soft-thresholding algorithm (ISTA). 
    We then develop two neural network architectures based on this algorithm: Convolutional ISTA-Net (CISTA-Net) and its enhanced version, CISTA-Net+.
    These networks transform the iterative ISTA process into a series of learnable layers, enabling end-to-end optimization of the CDL.
    \item To overcome the limitations of linear convolutional filters in CDL for capturing complex hybrid-field channel structures, we propose a CNN-CDL approach.
    This approach builds upon the PGD algorithm, enhancing it with deep learning components. 
    Specifically, we replace linear convolution filters in the gradient descent step of PGD with CNN blocks, and employ a learnable proximal mapping module in the proximal mapping step of PGD. 
    Additionally, we incorporate cross-layer feature integration in CNN-CDL to enhance feature flow across layers.
\end{itemize}

The remainder of this paper is organized as follows.
Section ~\ref{sec:system_model} introduces the system model and formulates the hybrid-field channel estimation problem.
Section~\ref{sec:3} formulates the CDL as a bilevel optimization problem and presents solution methods.
Section ~\ref{sec:4} presents the CNN-CDL in detail.
Sections~\ref{sec:5} and \ref{sec:6} provide simulation results and conclusion.

$\textit{Notations}$:
 Vectors and matrices are represented by boldface lowercase and uppercase letters, respectively. The transpose and conjugate transpose of a matrix $\mathbf{X}$ are denoted by $\mathbf{X}^T$ and $\mathbf{X}^H$, respectively. The $\ell_1$-norm and $\ell_2$-norm of a vector $\mathbf{x}$ are denoted by $\|\mathbf{x}\|_1$ and $\|\mathbf{x}\|_2$, respectively. $\operatorname{diag}(\mathbf{x})$ represents a diagonal matrix with the entries of vector $\mathbf{x}$ on its main diagonal. The Kronecker and Khatri-Rao products between two matrices $\mathbf{X}$ and $\mathbf{Y}$ are denoted by $\mathbf{X} \otimes \mathbf{Y}$ and $\mathbf{X} \diamond \mathbf{Y}$.
 
\section{SYSTEM MODEL AND PROBLEM FORMULATION}\label{sec:system_model}
\subsection{System Model}
\begin{figure} 
\centering 
\includegraphics{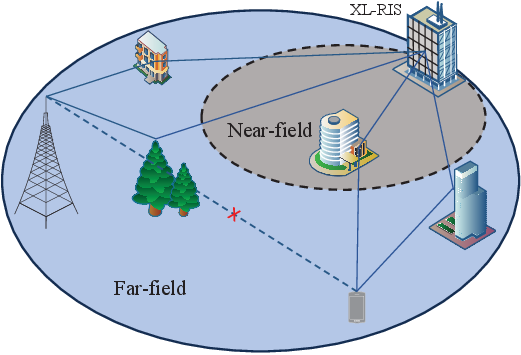} 
\caption{System model} 
\label{fig_1} 
\end{figure}

As illustrated in Fig.\ref{fig_1}, we consider an XL-RIS-aided mmWave massive MIMO system operating in time division duplex (TDD) mode, where a base station (BS) serves a single-antenna user\footnote{The results in this paper can be readily extended to the multiple users.} with the assistance of an XL-RIS.
In this system, the BS and XL-RIS are equipped with a uniform linear array (ULA) of $N$ antennas and a uniform planar array (UPA) of $M = M_1 \times M_2$ passive reflecting elements, respectively.
Since the direct communication link between the user and the BS is blocked by obstacles, the XL-RIS is installed on the building's outer surface, establishing a communication link between them.
The XL-RIS is connected to a controller that dynamically adjusts the phase of each reflecting elements to achieve high beamforming gain.
However, such a phase shift adjustment requires the CSI of the links between the BS, the XL-RIS, and the user.
To obtain the necessary CSI, the user sends pilot signals to the BS for uplink channel estimation.
The BS then utilizes the estimated channel for RIS phase shift adjustment and downlink data transmission, leveraging the channel reciprocity of TDD mode.
We adopt quasi-static block fading model, in which each channel remains approximately constant within a channel coherence block with $Q$ time slots.
  Assuming that $P$ slots of each coherence block are allocated for uplink channel estimation, and the remaining $Q-P$ time slots are used for downlink data transmission.
The uplink channel from the user to the RIS and from the RIS to the BS are denoted by $\mathbf{h}_r \in \mathbb{C}^{M \times 1}$ and  $\mathbf{G} \in \mathbb{C}^{N \times M}$, respectively.
The phase shift vector of RIS is denoted by $\boldsymbol{\theta}=\left[e^{j \theta_1}, \cdots, e^{j \theta_M}\right]^{{T}} \in \mathbb{C}^{M \times 1}$, where $ e^{j \theta_m} $ is the coefficient of the $m$-th reflecting element.
At time slot $t$, the received pilot signal at the BS can be expressed as
\begin{equation}
\begin{aligned}
\mathbf{y}_t & =\mathbf{G}\operatorname{diag}\left(\boldsymbol{\theta}_t\right) \mathbf{h}_r s_t+\mathbf{n}_t\\
&=\mathbf{G}\operatorname{diag}\left(\mathbf{h}_r\right) \boldsymbol{\theta}_t s_t+\mathbf{n}_t,
\label{eq_1} 
\end{aligned}
\end{equation}
where $s_t$ is the transmitted pilot signal and $\mathbf{n}_t\sim\mathcal{C N}\left(\mathbf{0}, \sigma^2 \mathbf{I}\right)$ is the Gaussian noise vector with the noise power $\sigma^2$.
For simplicity, the value of $s_t$ is set to 1, as the pilot signal is perfectly known to the BS. The effective cascaded channel in (\ref{eq_1}) can be defined as
\begin{equation}
    \mathbf{H} = \mathbf{G}\operatorname{diag}\left(\mathbf{h}_r\right).
    \label{eq_2}
\end{equation}

After receiving $P$ successive pilot signal, the received signal matrix $\mathbf{Y}=\left[\mathbf{y}_1, \ldots, \mathbf{y}_{P}\right] \in \mathbb{C}^{N \times P} $ at the BS can be written as
\begin{equation}
\mathbf{Y}= \mathbf{H} \mathbf{\Theta}+\mathbf{N},
\label{eq_3} 
\end{equation}
where $\mathbf{\Theta}=\left[\boldsymbol{\theta}_1, \cdots, \boldsymbol{\theta}_P\right]$ is the phase shift matrix and $\mathbf{N}=\left[\mathbf{n}_1, \ldots, \mathbf{n}_{P}\right]$ is the noise matrix.

\subsection{Channel Model}
We start by discussing the far-field and near-field channel models, which form the basis for introducing the hybrid-field channel model for XL-RIS-aided massive MIMO systems.
The Rayleigh distance $R$, a commonly used measure to distinguish between the far-field and near-field regions, can be defined as
\begin{equation}
R=\frac{2 D^2}{\lambda},
\label{eq_5}
\end{equation}
where $D$ is the array aperture and $\lambda$ is the wavelength.
If the distance between the array and receiver (or scatter) exceeds the Rayleigh distance, the radiation field is in the far-field region. Otherwise, the radiation field is in the near-field region. 
For instance, considering an XL-RIS with array aperture $D=0.9$ m, operating at 10 GHz with a carrier wavelength of 0.03 m, the Rayleigh distance is $54$ m.
In this case, some scatterers are likely to be in the near-field region, while others are in the far-field region, resulting in a hybrid-field scenario.

In the far-field region, the propagation wave can be well approximated as a planar wavefront.
For the planar wave model, the approximation that the signals impinge on all array elements with the same direction of incidence is made to simplify the array steering vector. The array steering vector of ULA at BS\cite{9919846} is defined as
\begin{equation}
\mathbf{a}(\psi)=\frac{1}{\sqrt{N}}\left[ e^{j \frac{2 \pi d}{\lambda} sin(\psi)\mathbf{n}}\right]^T,
\label{eq_6}
\end{equation}
where $\mathbf{n}=\left[0,1, \cdots, N-1\right]$, $\psi $ is the physical angle of arrival, and $d$ is array element spacing, satisfying $d=\lambda / 2$.
The array steering vector of UPA at XL-RIS is defined as
\begin{equation}
\boldsymbol{b}_{f a r}\left(\phi, \varphi \right)=\boldsymbol{b}_{v}(\phi) \otimes \boldsymbol{b}_{h}(\phi, \varphi)  ,
\label{eq_7}
\end{equation}
where $\phi, \varphi$ are the elevation and azimuth angles of departure, respectively.
The steering vectors $\boldsymbol{b}_{v}(\phi)$ and $\boldsymbol{b}_{h}(\phi, \varphi)$ correspond to the vertical and horizontal directions of the UPA. These vectors are defined as\cite{9919846}

\begin{equation}
\begin{aligned}
\boldsymbol{b}_{v}(\phi) &=\frac{1}{\sqrt{M_1}}\left[ e^{j \frac{2 d \pi}{\lambda} cos(\phi)\mathbf{m_1}}\right]^T,\\
\boldsymbol{b}_{h}(\phi, \varphi) &= \frac{1}{\sqrt{M_2}} \left[ e^{j \frac{2 d \pi}{\lambda} \sin(\phi) \cos(\varphi)\mathbf{m_2}}\right]^T,
\end{aligned}
\label{eq_8}
\end{equation}
where $\mathbf{m_1}=\left[0,1, \cdots, M_1-1\right]$ and $\mathbf{m_2}=\left[0,1, \cdots, M_2-1\right]$.

As illustrated in Fig.\ref{fig_1}, the BS is located in the far-field region of the XL-RIS.
Given this configuration and the limited scattering characteristics of the environment, the channel between the XL-RIS and the BS can be described using the Saleh-Valenzuela (SV) model\cite{6717211}
\begin{equation}
\mathbf{G}=\sum_{i=1}^{L_1} \alpha_i \mathbf{a}(\psi_i) \boldsymbol{b}^{H}_{f a r}\left(\phi_i, \varphi_i \right),
\label{eq_9}
\end{equation}
where $L_1$ is the number of propagation paths in RIS-BS link, $\alpha_i$ is the complex gain of the $i$-th path, and $\left(\phi_i, \varphi_i \right) $ and $\psi_i$ are the AoD and AoA of the $i$-th path, respectively. 
For a more concise representation, we can reformulate equation \eqref{eq_9} as
\begin{equation}
\mathbf{G}=\mathbf{A}\left(\boldsymbol{\psi}_A\right) \boldsymbol{\Lambda}_{\alpha} \mathbf{B}^{H}_{f a r}\left(\boldsymbol{\phi}_D, \boldsymbol{\varphi}_D\right),
\label{eq_11}
\end{equation}
where $\boldsymbol{\Lambda}_{\alpha}=\operatorname{diag}\left(\alpha_1, \cdots, \alpha_{L_1}\right)$ is the path gain matrix, $\mathbf{A}\left(\boldsymbol{\psi}_A\right)=\left[\mathbf{a}(\psi_1), \ldots, \mathbf{a}(\psi_{L_1})\right]$ is the AoA steering matrix, and $\mathbf{B}_{f a r}\left(\boldsymbol{\phi}_D,\boldsymbol{\varphi}_D\right)=\left[\boldsymbol{b}_{f a r}\left(\phi_1, \varphi_1 \right), \ldots, \boldsymbol{b}_{f a r}\left(\phi_{L_1}, \varphi_{L_1} \right)\right]$ is the AoD steering matrix.

In this paper, we consider a realistic hybrid-field communication scenario between RIS and user, as shown in Fig.\ref{fig_1}.
The channel in this scenario comprises both far-field and near-field path components, each arising from their corresponding scatterers.
We first introduce the far-field channel, which can be modeled similarly to the BS-RIS channel
\begin{equation}
\mathbf{h}_{f a r}=\sum_{j=1}^{L_f} \beta_j \boldsymbol{b}_{f a r}\left(\phi_j, \varphi_j \right),
\label{eq_10}
\end{equation}
where $L_f$ is the number of far-field propagation paths in RIS-user link, $\beta_j$ is the complex gain of the $j$-th path, and $\left(\phi_j, \varphi_j \right)$ is the AOA of the $j$-th path.

For the near-field channel, the propagation wave can be more accurately represented as a spherical wavefront than a planar wavefront.
For the spherical wave model, the array steering vector depends on both the incident angle and the distances between the reflecting elements of the RIS and the scatterers.
The near-field steering vector at the XL-RIS \cite{10123941} is given by
\begin{equation}
\begin{array}{r}
\mathbf{b}_{n e a r}\left(\mathbf{r}\right)=\frac{1}{\sqrt{M}}\left[e^{-j \frac{2 \pi}{\lambda} r(0,0)}, \ldots, e^{-j \frac{2 \pi}{\lambda} r\left(0, M_2-1\right)}, \ldots,\right. \\
\left.e^{-j \frac{2 \pi}{\lambda} r\left(M_1-1, 0\right)}, \ldots, e^{-j \frac{2 \pi}{\lambda} r\left(M_1-1, M_2-1\right)}\right]^T,
\end{array}
\label{eq_12}
\end{equation}
where $\mathbf{r}=\left[ r(0,0), \ldots, r\left(M_1-1, M_2-1\right)\right]^T$  is the distance vector and $r\left(m_1, m_2\right)$ denote the distance from scatterer to the $\left(m_1, m_2\right)$-th element of the XL-RIS.
Similar to \eqref{eq_10}, the RIS-user channel in the near-field region can be formulated as
\begin{equation}
\mathbf{h}_{n e a r}=\sum_{j=1}^{L_n} \beta_j \boldsymbol{b}_{near}\left(\mathbf{r}_j \right).
\label{eq_13}
\end{equation}

Having introduced both the far-field and near-field channel separately, the overall RIS-user channel in this hybrid-field scenario can be expressed as
\begin{equation}
\begin{aligned}
\mathbf{h}_{hybrid}&=\sum_{j=1}^{L_f} \beta_j \boldsymbol{b}_{f a r}\left(\phi_j, \varphi_j \right)+\sum_{j=1}^{L_n} \beta_j \boldsymbol{b}_{near}\left(\mathbf{r}_j \right)\\
&=\mathbf{B}\left(\boldsymbol{r},\boldsymbol{\phi}_A, \boldsymbol{\varphi}_A\right)\boldsymbol{\beta},
\end{aligned}
\label{eq_14}
\end{equation}
where $\mathbf{B}\left(\boldsymbol{r},\boldsymbol{\phi}_A, \boldsymbol{\varphi}_A\right)=\left[\boldsymbol{b}\left(\mathbf{r}_1, \phi_1, \varphi_1 \right), \cdots, \boldsymbol{b}\left(\mathbf{r}_{L_2}, \phi_{L_2}, \varphi_{L_2} \right)\right]$  and $\boldsymbol{\beta}=\left[\beta_1, \cdots, \beta_{L_2}\right]^{T}$ is the path gain vector, with $L_2=L_n+L_f$ representing the total number of paths in both near-field and far-field components.
The expression of $\boldsymbol{b}\left(\mathbf{r}_j, \phi_j, \varphi_j \right)$ depends on whether the distance $r_j=\mathbf{r}_j[0]$ between the $j$-th scatterer and RIS is greater than or less than the Rayleigh distance $R$
\begin{equation}
\boldsymbol{b}\left(\mathbf{r}_j, \phi_j, \varphi_j \right)= \begin{cases}\boldsymbol{b}_{f a r}\left(\phi_j, \varphi_j \right), & \text { if } r_j>R \\ \mathbf{b}_{n e a r}\left(\mathbf{r}_j\right), & \text { otherwise. }\end{cases}
\label{eq_15}
\end{equation}

Substituting $\mathbf{G}$ and $\mathbf{h}_{hybrid}$ in \eqref{eq_11} and \eqref{eq_14} into the cascaded channel $\mathbf{H}$ in \eqref{eq_2}, we obtain the following reformulated cascaded channel
\begin{equation}
\begin{aligned}
\mathbf{H} & = \mathbf{G}\operatorname{diag}\left(\mathbf{h}\right) {=} \mathbf{G} \diamond \mathbf{h}^{T} \\
& \stackrel{}{=}\left(\mathbf{A}\boldsymbol{\Lambda}_{\alpha} \mathbf{B}_{f a r}^{{H}}\right) \diamond \left(\mathbf{B}\boldsymbol{\beta}\right)^T \\
& \stackrel{(a)}{=}\left(\mathbf{A}\boldsymbol{\Lambda}_{\alpha} \otimes\boldsymbol{\beta}^{T}\right) \left(\mathbf{B}_{f a r}^{H}\diamond \mathbf{B}^{T}\right)\\
& \stackrel{(b)}{=}\mathbf{A}\left(\boldsymbol{\Lambda}_{\alpha} \otimes \boldsymbol{\beta}^{T}\right)\left(\mathbf{B}_{f a r}^{H}\diamond \mathbf{B}^{T}\right)\\
\end{aligned}
\label{eq_16}
\end{equation}
where $(a)$ follows from the property of Khatri-Rao product and $(b)$ relies on the property of Kronecker product.
The signal model in \eqref{eq_3} can be reformulated into a more compact vector form by vectorizing the $\mathbf{Y}$
\begin{equation}
\begin{aligned}
\mathbf{y} & = \operatorname{vec}\left( \mathbf{H} \mathbf{\Theta}\right)  +\operatorname{vec}\left(\mathbf{N}\right)\\
& \stackrel{a}{=}\left(\mathbf{\Theta}^{T}\otimes\mathbf{I}_N\right)\operatorname{vec}\left(\mathbf{H}\right)  +\boldsymbol{n} \\
& \stackrel{(b)}{=}\left(\mathbf{\Theta}^{T}\otimes\mathbf{I}_N\right)\operatorname{vec}\left(\mathbf{A}\left(\boldsymbol{\Lambda}_{\alpha} \otimes \boldsymbol{\beta}^{T}\right)\left(\mathbf{B}_{f a r}^{H}\diamond \mathbf{B}^{T}\right)\right)  +\boldsymbol{n}\\
& \stackrel{(c)}{=}\left(\mathbf{\Theta}^{T}\otimes\mathbf{I}_N\right)\left(\left(\mathbf{B}_{f a r}^{H}\diamond \mathbf{B}^{T}\right)^{T}\otimes\mathbf{A}\right)\operatorname{vec}\left(\boldsymbol{\Lambda}_{\alpha} \otimes \boldsymbol{\beta}^{T}\right)  +\boldsymbol{n}\\
&\stackrel{(d)}{=} \boldsymbol{\Phi} \mathbf{D}\boldsymbol{g}+\boldsymbol{n},
\end{aligned}
\label{eq_17}
\end{equation}
where $(a)$ and $(c)$ are based on the property of Kronecker product, $\boldsymbol{\Phi} \triangleq\mathbf{\Theta}^{T}\otimes\mathbf{I}_N$ is the measurement matrix, $\mathbf{D}\triangleq \left(\mathbf{B}_{f a r}^{H}\diamond \mathbf{B}^{T}\right)^{T}\otimes\mathbf{A}$ is the dictionary matrix, and $\boldsymbol{g} = \operatorname{vec}\left(\boldsymbol{\Lambda}_{\alpha} \otimes \boldsymbol{\beta}^{T}\right)$ is the path gain vector.

\subsection{Problem Formulation}
The vectorized representation in \eqref{eq_17} clearly illustrates the channel estimation challenges. Specifically, it highlights two key unknowns that need to be determined: the dictionary matrix $\mathbf{D}$ and the path gain vector $\boldsymbol{g}$, both crucial for accurate channel estimation.

Compressive sensing provides an effective framework for channel estimation by leveraging the sparsity of the channel in an overcomplete DFT dictionary.
This framework has been successfully applied in far-field channel estimation, where sparsity in the angular domain can be exploited with a predefined overcomplete dictionary.
Similarly, for near-field channels, sparsity in the polar domain can also be utilized.
However, for XL-RIS cascaded channels in hybrid-field scenarios, the combination of hybrid-field propagation and the cascaded nature of these channels makes it difficult to define an appropriate overcomplete dictionary in advance.

To address this challenge, we formulate the channel estimation problem as a dictionary learning problem, aiming to obtain $\mathbf{D}$ and $\boldsymbol{g}$ by solving the following optimization problem
{\begin{equation}
\min _{\mathbf{D},\boldsymbol{g} }\frac{1}{2}\|\mathbf{y}-\boldsymbol{\Phi} \mathbf{D}\boldsymbol{g}\|_2^2+\rho R(\boldsymbol{g}),
\label{eq_18}
\end{equation}
where $R(\cdot)$ is a regularization term that promotes desirable properties of $\boldsymbol{g}$, such as sparsity. 
The parameter $\rho$ controls the balance between the data fidelity term $\|\mathbf{y}-\boldsymbol{\Phi} \mathbf{D}\boldsymbol{g}\|_2^2$ and the regularization term.

\section{Dictionary Learning for Hybrid-Field Channels}\label{sec:3}
\subsection{Dictionary Learning}
Dictionary learning is a signal representation framework that aims to find an optimal set of basis vectors, termed atoms, that can effectively represent a given class of signals\cite{5714407}.
In the context of channel estimation, dictionary learning enables expressing the channel as a linear combination of atoms ${\mathbf{d}_s}$ from a learned dictionary $\mathbf{D}=\left\{\mathbf{d}_1, \mathbf{d}_2, \ldots, \mathbf{d}_S\right\}$, which can be formulated as
\begin{equation}
    \mathbf{h}=\sum_{s=1}^S g_{s} \mathbf{d}_{s}=\mathbf{D}\boldsymbol{g} ,
    \label{eq_19}
\end{equation}
where $\boldsymbol{g} = [g_1, g_2, \ldots, g_{S}]^T$ is the vector of coefficients corresponding to each atom in the dictionary.

Dictionary learning offers a flexible framework to overcome the limitations of predefined dictionaries. 
Learning the dictionary matrix $\mathbf{D}$ from the received pilot signal enables a more effective capture of the characteristics in the channel.
Conventional dictionary learning algorithms, such as K-SVD\cite{1710377}, have been successfully applied to various low-dimensional signal processing tasks. 
However, when dealing with high-dimensional signals, these methods often employ a patch-based strategy\cite{5466111}, wherein the signal is divided into patches of smaller dimensions, and the dictionary is learned on these low-dimensional patches. 
In XL-RIS systems, the channel dimension can be significantly large. 
Directly applying conventional dictionary learning methods to such high-dimensional channel is computationally prohibitive and might lead to numerical instabilities. 
Nonetheless, the patch-based strategy is unsuitable for channel estimation, as dividing the channel into patches would violate its inherent structure.

\subsection{Convolutional Dictionary Learning}
To address the challenges posed by high-dimensional channel in XL-RIS systems, we propose a convolutional dictionary learning approach\cite{8364626} .
CDL extends the conventional dictionary learning framework by efficiently handling high-dimensional signals through the use of convolutional operations instead of matrix multiplication.
This approach leverages the shift-invariant property of convolutions to effectively capture the inherent structural patterns within the entire high-dimensional channel.
The CDL approach for representing the channel can be formulated as
\begin{equation}
\mathbf{h}  = \sum_{s=1}^S \bar{\mathbf{d}}_{s} \ast \bar{\boldsymbol{g}}_{s} = \bar{\mathbf{D}}\bar{\boldsymbol{g}},
\label{eq_21}
\end{equation}
where $\bar{\mathbf{d}}_{s}$ is the convolutional filters, $\ast$ denotes the convolutional operator, and $\bar{\boldsymbol{g}}_{s}$ is the  corresponding coefficient maps.
For ease of expression and leveraging the properties of convolution, this formulation can be written in matrix form\cite{plaut2019greedy}, in which $\bar{\boldsymbol{g}}=\left[\bar{\boldsymbol{g}}_{1}, \cdots, \bar{\boldsymbol{g}}_{S}\right]$ and $\bar{\mathbf{D}}=\left[\bar{\mathbf{D}}_{1}, \cdots, \bar{\mathbf{D}}_{S}\right]$, with $\bar{\mathbf{D}}_{s}$ being a Toeplitz matrix corresponding to $\bar{\mathbf{d}}_{s}$.

Building upon this CDL representation, the hybrid-field channel estimation problem can be formulated as
\begin{equation}
\min _{\bar{\mathbf{D}},\bar{\boldsymbol{g}} }\frac{1}{2}\|\mathbf{y}-\boldsymbol{\Phi} \bar{\mathbf{D}}\bar{\boldsymbol{g}}\|_2^2+\rho R(\bar{\boldsymbol{g}}).
\label{eq_20}
\end{equation}
The optimization problem in \eqref{eq_20} is challenging due to its non-convexity and the interdependence between the dictionary $\bar{\mathbf{D}}$ and the coefficient maps $\bar{\boldsymbol{g}}$. 
To address these challenges, we propose a bilevel optimization framework\cite{10502023} for CDL. 
The bilevel optimization formulation can be expressed as
\begin{subequations}
\begin{align}
& \min_{\bar{\mathbf{D}}} F(\bar{\mathbf{D}}, \bar{\boldsymbol{g}}^*(\bar{\mathbf{D}})) = \min_{\bar{\mathbf{D}} }   \|\mathbf{y} - \boldsymbol{\Phi} \bar{\mathbf{D}} \bar{\boldsymbol{g}}^*(\bar{\mathbf{D}})\|_2^2 
\label{eq:bilevel_optimization_compact_a} \\
& \text{s.t.} \quad \bar{\boldsymbol{g}}^*(\bar{\mathbf{D}}) = \arg\min_{\bar{\boldsymbol{g}}} \frac{1}{2}\|\mathbf{y} - \boldsymbol{\Phi} \bar{\mathbf{D}} \bar{\boldsymbol{g}}\|_2^2 + \rho R(\bar{\boldsymbol{g}}),
\label{eq:bilevel_optimization_compact_b}
\end{align}
\end{subequations}
where $\bar{\boldsymbol{g}}^*(\bar{\mathbf{D}})$ denotes the optimal coefficient maps given $\bar{\mathbf{D}}$, obtained by solving the lower-level problem \eqref{eq:bilevel_optimization_compact_b}. 
The upper-level optimization \eqref{eq:bilevel_optimization_compact_a} aims to identify the dictionary $\bar{\mathbf{D}}$ minimizing the reconstruction error.

To tackle this bilevel optimization problem, we adopt a gradient-based approach, where the gradient of the upper-level objective $F$ with respect to the dictionary $\bar{\mathbf{D}}$ is computed via the chain rule as follows
\begin{equation}
\nabla F(\bar{\mathbf{D}}) = \nabla_{\bar{\mathbf{D}}} F(\bar{\mathbf{D}}, \bar{\boldsymbol{g}}^*(\bar{\mathbf{D}})) +  \left(\nabla_{\bar{\mathbf{D}}} \bar{\boldsymbol{g}}^*(\bar{\mathbf{D}})\right)^{T} \nabla_{\bar{\boldsymbol{g}}^*} F(\bar{\mathbf{D}}, \bar{\boldsymbol{g}}^*(\bar{\mathbf{D}}))
  \end{equation}
However, obtaining the term $ \nabla_{\bar{\mathbf{D}}} \bar{\boldsymbol{g}}^*(\bar{\mathbf{D}})$ poses a significant challenge, which arises from the lack of an explicit expression for $\bar{\boldsymbol{g}}^*(\bar{\mathbf{D}})$ due to the difficulty in solving the lower-level problem.

To overcome this challenge, we propose an iterative approximation approach.
This approach approximates the solution to the lower-level problem by employing an iterative optimization algorithm.
At each iteration $k \in [1, \ldots, K]$, the approximate solution $\bar{\boldsymbol{g}}_{k}$ is updated according to 
\begin{equation}
\bar{\boldsymbol{g}}_{k} = \boldsymbol{\Psi}(\bar{\boldsymbol{g}}_{k-1}, \bar{\mathbf{D}}),
\end{equation}
where $\boldsymbol{\Psi}$ represents the iterative update rule.
Through this iterative approximation of the lower-level optimization, the bilevel problem can be reformulated as
\begin{subequations}
\begin{align}
& \min_{\bar{\mathbf{D}}} F(\bar{\mathbf{D}}, \bar{\boldsymbol{g}}_{K}(\bar{\mathbf{D}})) = \min_{\bar{\mathbf{D}} }   \|\mathbf{y} - \boldsymbol{\Phi} \bar{\mathbf{D}} \bar{\boldsymbol{g}}_{K}(\bar{\mathbf{D}})\|_2^2 \label{eq:unroll_bilevel1} \\
& \text{s.t.} \quad\bar{\boldsymbol{g}}_{k} = \boldsymbol{\Psi}(\bar{\boldsymbol{g}}_{k-1}, \bar{\mathbf{D}}),  \quad \forall k \in[1 \ldots K].
\label{eq:unroll_bilevel2}
\end{align}
\end{subequations}
This reformulation facilitates gradient computation with respect to $\bar{\mathbf{D}}$ by enabling differentiation through the iterative process. 
The resulting gradient expression is given as
\begin{equation}
\begin{aligned}
\nabla F(\bar{\mathbf{D}}) &= \nabla_{\bar{\mathbf{D}}} F(\bar{\mathbf{D}}, \bar{\boldsymbol{g}}_K(\bar{\mathbf{D}}))  \\
&\quad + \left( \sum_{k=1}^K \left( {H}_K \cdots {H}_{k+1} \right) {J}_k \right)^{T} \nabla_{\bar{\boldsymbol{g}}} F(\bar{\mathbf{D}}, \bar{\boldsymbol{g}}_K(\bar{\mathbf{D}})),
\end{aligned}
\label{eq_gri_up}
\end{equation}
where
$ H_k =\nabla_{\bar{\boldsymbol{g}}} \boldsymbol{\Psi}(\bar{\boldsymbol{g}}_{k-1}, \bar{\mathbf{D}})$  and $J_k =\nabla_{\bar{\mathbf{D}}} \boldsymbol{\Psi}(\bar{\boldsymbol{g}}_{k-1}, \bar{\mathbf{D}})$.
 \subsection{Proximal Gradient Descent for Lower-Level Problem}
The approximation approach for the lower-level problem in the bilevel optimization requires an efficient algorithm to update the coefficient maps $\bar{\boldsymbol{g}}$ at each iteration. 
To address this requirement, we employ the proximal gradient descent algorithm \cite{parikh2014proximal} as our iterative optimization algorithm.
PGD is an effective algorithm for solving non-smooth convex problems, making it particularly suitable for the lower-level problem.

As an iterative algorithm, PGD performs updates at each iteration according to the following rules
\begin{align}
\boldsymbol{z}_{k} & = \bar{\boldsymbol{g}}_{k-1} + \lambda\bar{\mathbf{D}}^{T} \boldsymbol{\Phi}^T \left(\mathbf{y} - \boldsymbol{\Phi} \bar{\mathbf{D}} \bar{\boldsymbol{g}}_{k-1}\right), \label{eq_z_update} \\
\bar{\boldsymbol{g}}_{k} & = \text{prox}_{\rho, R}(\boldsymbol{z}_{k}), \label{eq_g_update}
\end{align}
where $\lambda$ denotes the step size, and $\text{prox}_{\rho, R}(\cdot)$ represents the proximal mapping associated with the regularization term $R(\cdot)$. The proximal mapping is defined as
\begin{equation}
\text{prox}_{\rho, R}(\boldsymbol{z}_{k}) = \underset{\bar{\boldsymbol{g}}}{\arg \min} \frac{1}{2}\left\|\bar{\boldsymbol{g}} - \boldsymbol{z}_{k}\right\|_2^2 + \rho R(\bar{\boldsymbol{g}}).
\end{equation}

When the regularization term $R(\cdot)$ is chosen to be the $\ell_1$ norm, i.e., $R(\bar{\boldsymbol{g}}) = \|\bar{\boldsymbol{g}}\|_1$, the proximal mapping simplifies to the well-known soft-thresholding operator.
In this case, the PGM algorithm reduces to the iterative soft-thresholding algorithm \cite{beck2009fast}. 
The  $\bar{\boldsymbol{g}}_{k}$ is updated using the soft-thresholding operator $S_{\rho, \lambda}(\cdot)$ as follows
\begin{equation}
\bar{\boldsymbol{g}}_{k}=S_{\rho, \lambda}\left(\boldsymbol{z}_{k-1}\right),
\label{eq_ista}
\end{equation} 
where $S_{\rho, \lambda}(\cdot)$ is defined component-wise for each element $z$ in $\boldsymbol{z}$ by
\begin{equation}
S_{\rho, \lambda}(z) = \max \left\{\left|z\right|-\rho \lambda, 0\right\} \operatorname{sign}\left(z\right),
\end{equation}
The soft-thresholding operator effectively shrinks the components of its input towards zero, which introduces sparsity into the solution of the optimization problem. 

\subsection{Convolutional ISTA-Net}
Building upon the ISTA algorithm for addressing the lower-level problem with an $\ell_1$ regularizer, we transform it into a structured neural network architecture through algorithm unrolling\cite{9363511}. 
This transformation yields the convolutional ISTA network (CISTA-Net). 
CISTA-Net combines the advantages of deep learning with the interpretability of the ISTA.

In the CISTA-Net, each layer corresponds to one iteration of the ISTA.
To align with the nature of convolutional dictionary learning, we replace the matrix multiplications involving $\bar{\mathbf{D}}$ in \eqref{eq_z_update} with learnable convolutional kernels $\mathbf{D}^{c}$. 
Specifically, in the $k$-th layer of CISTA-Net, the computation of $\boldsymbol{z}_{k}$ is performed through the following operations
\begin{equation}
\boldsymbol{z}_{k} = \bar{\boldsymbol{g}}_{k-1} + \lambda \mathbf{D}^{c} \circledast \left(\boldsymbol{\Phi}^T \left( \mathbf{y} - \boldsymbol{\Phi}\left( \mathbf{D}^{c} \ast \bar{\boldsymbol{g}}_{k-1} \right)\right)\right), \label{eq_z_update_conv_Dsub}
\end{equation}
where $\circledast$ is the transposed convolution operator.

The forward pass of this network computes the solution to the lower-level problem \eqref{eq:unroll_bilevel2}.
In the backward pass, the loss function is defined as the objective of the upper-level problem \eqref{eq:unroll_bilevel1}.
By leveraging automatic differentiation in deep learning libraries like PyTorch, the gradients of the loss function with respect to the convolutional kernels $\mathbf{D}^{c}$ can be efficiently computed.
The automatic differentiation mechanism eliminates the need for manually implementing the complex expression in \eqref{eq_gri_up} during the optimization process.
These gradients are then used in conjunction with advanced optimization algorithms like Adam to update the convolutional kernels and minimize the upper-level problem's loss.
By unrolling ISTA into a network with learnable convolutional kernels, the CISTA-Net enables an end-to-end manner to solve the bilevel optimization problem.

\subsection{Convolutional ISTA-Net+}
CISTA-Net integrates the ISTA algorithm into a neural network architecture and allows for the learning of a convolutional dictionary. 
However, it keeps the parameters $\lambda$, $\rho$, and $\boldsymbol{\Phi}$ fixed during training.
To enhance the network's adaptability and performance, we introduce CISTA-Net+, which incorporates additional learnable parameters.

In CISTA-Net+, each layer contains its own set of learnable parameters $\lambda_k$, $\rho_k$, and decoupled measurement matrices $\boldsymbol{B}_{k}$ and $\boldsymbol{C}_{k}$.
Additionally, the network employs layer-specific convolutional dictionaries $\mathbf{A}_{k}^{c}$ and $\mathbf{D}_{k}^{c}$, allowing it to capture multi-layer features.
This design significantly enhances the network's representational capacity compared to conventional CDL methods, which typically employ a single dictionary throughout the entire process.
The update for $\boldsymbol{z}_{k}$ in CISTA-Net+ is expressed as
\begin{equation}
\boldsymbol{z}_{k} = \bar{\boldsymbol{g}}_{k-1} + \lambda_{k} \mathbf{A}_{k}^{c} \circledast \left(\boldsymbol{B}_{k} \left( \mathbf{y} - \boldsymbol{C}_{k}\left( \mathbf{D}_{k}^{c} \ast \bar{\boldsymbol{g}}_{k-1} \right)\right)\right).
\label{eq_z_update_conv_Dsub+}
\end{equation}
Furthermore, the soft-thresholding operator incorporates a learnable threshold $\rho_k$:
\begin{equation}
\bar{\boldsymbol{g}}_{k}=S_{\rho_k, \lambda_k}\left(\boldsymbol{z}_{k}\right).
\label{eq_cista+}
\end{equation} 

After processing through all $K$ layers, the network's output $\bar{\boldsymbol{g}}_{K}$ is convolved with an additional dictionary $\mathbf{D}_{K+1}^{c}$ to obtain the reconstructed channel.
CISTA-Net+ adopts an end-to-end training approach, optimizing all learnable parameters simultaneously.
Unlike CISTA-Net, which employs a self-supervised learning strategy with the loss function defined as the objective of the upper-level problem in \eqref{eq:unroll_bilevel1}, the CISTA-Net+ adopts a supervised learning strategy.
The loss function is defined as the mean squared error between the reconstructed channel and the true channel
   \begin{equation}
\mathcal{L}_{\text{CISTA-Net+}} =  \|\mathbf{h} - \mathbf{D}_{K+1}^{c} \ast \bar{\boldsymbol{g}}_{K} \|_2^2.
\label{eq_loss_cistanet+}
\end{equation}

\section{Deep Dictionary Learning for Hybrid-Field Channels}\label{sec:4}
\subsection{CNN-CDL Framework}
In the previous section, CDL-based methods such as CISTA-Net and CISTA-Net+ demonstrated considerable promise in addressing the challenges of high-dimensional channel estimation. 
These methods leverage convolutional dictionaries to capture inherent patterns in high-dimensional channels. 
However, the linear nature of these dictionaries might limit their ability to fully capture the complex structures present in hybrid-field channel.

To further enhance the accuracy of hybrid-field channel
estimation, we propose an approach called CNN-CDL, which
integrates convolutional neural network (CNN) with CDL.
CNN is widely recognized for its modeling capabilities and feature extraction.
These characteristics make CNN an ideal choice for extending CDL, enabling the CNN-CDL to better capture the 
intricate and non-linear structures within hybrid-field channel.

CNN-CDL builds on the foundations of the PGD algorithm, previously introduced as an effective optimization method for solving lower-level problems. 
The CNN-CDL unrolls the iterative optimization process of PGD, mapping each iteration into a specific layer within the network architecture, which in turn enables the integration of CNN-based components at each layer.
Each layer in the CNN-CDL comprises two primary steps: the gradient descent step as shown in \eqref{eq_z_update} and the proximal mapping step as shown in \eqref{eq_g_update}.
In the following subsections, we will detail the implementation of these steps.

\subsection{The Gradient Descent Step}
The gradient descent step in PGD employs convolutional operations with the dictionary. 
To address the limitation of the dictionary's linear nature in capturing complex structures of channel, we replace these operations with CNN blocks.

Specifically, two CNN blocks, denoted as $\mathcal{F}_{k}^{A}$ and $\mathcal{F}_{k}^{{{D}}}$, are introduced at each layer to take on the roles of the dictionary $\bar{\mathbf{D}}$ and its transpose, respectively.
Each block is constructed using a residual architecture that includes two convolutional layers followed by a PReLU non-linear activation function.
The modified gradient descent step in CNN-CDL, referred to as the gradient descent module (GDM), is expressed as
\begin{equation}
\boldsymbol{z}_{k} = \bar{\boldsymbol{g}}_{k-1} + \lambda_{k} \mathcal{F}_{k}^{A} \left(\boldsymbol{B}_{k} \left( \mathbf{y} - \boldsymbol{C}_{k} \mathcal{F}_{k}^{D}  \left(\bar{\boldsymbol{g}}_{k-1}\right) \right)\right). \label{eq_z_update_conv_Dsub++}
\end{equation}

By incorporating these CNN blocks in the gradient descent step, CNN-CDL enhances its ability to capture the intricate and nonlinear structures inherent in hybrid-field channels, thereby overcoming the limitations of a linear convolutional dictionary.

\subsection{The Proximal Mapping Step}
In the PGD algorithm, the proximal mapping step refines the solution by applying regularization to promote specific properties.
This step is guided by a regularization term, denoted as $R(\cdot)$, which imposes prior knowledge onto the solution.
Therefore, the choice of $R(\cdot)$ is crucial, as it significantly impacts the overall performance of the algorithm.
Conventional CDL methods assume specific properties of $\bar{\boldsymbol{g}}$ and then design the regularization function to enforce these assumptions. 
A common instance of $R(\cdot)$  is the $\ell_1$ norm, which is used to promote sparsity in the solution.
The proximal mapping based on this regularization term is implemented using the soft-thresholding operator, as shown in \eqref{eq_ista}.
\begin{figure*}[htbp]
        \centering
	{\includegraphics[scale=0.85]{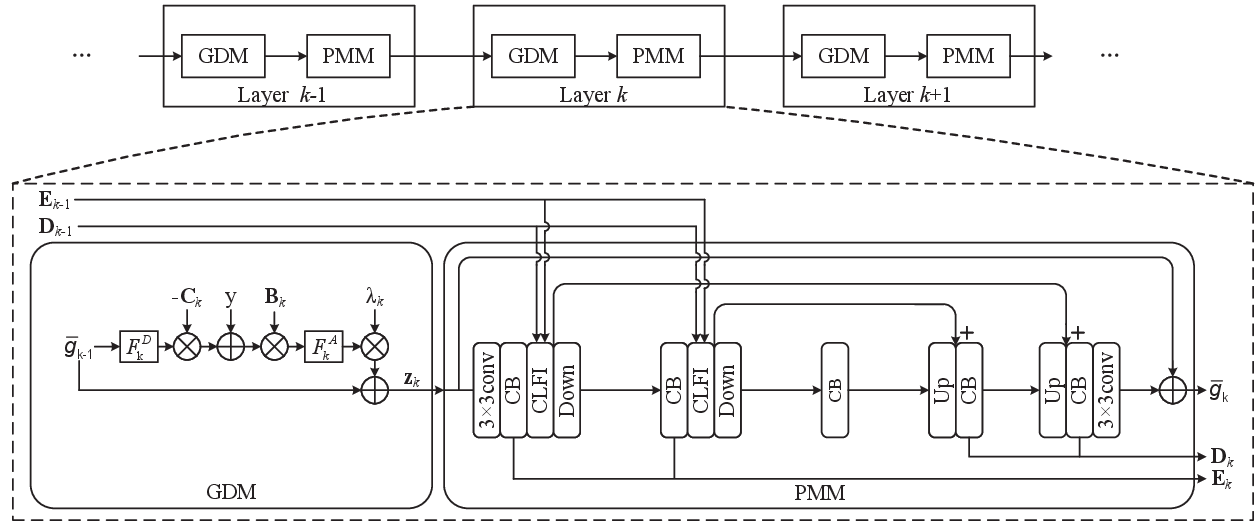}}
	\caption{Architecture of the proposed CNN-CDL network.}
        \label{CNN_CDL}
\end{figure*}

However, for hybrid-field cascaded channels, the complex nature of the channel makes it challenging to accurately define the prior knowledge of $\bar{\boldsymbol{g}}$.
Additionally, replacing linear convolutional dictionaries with CNN blocks further complicates the definition of prior knowledge due to the nonlinearity introduced by CNN blocks.
As a result, $\bar{\boldsymbol{g}}$ might have complex and unpredictable structures that cannot be easily captured by hand-crafted regularization functions due to the lack of suitable prior knowledge.
Consequently, without well-designed regularization functions, it becomes challenging to explicitly perform the proximal mapping step. 

To address these challenges, we propose a data-driven approach for the proximal mapping step.
This approach introduces a learnable operator that adapts to the complex structures of $\bar{\boldsymbol{g}}$, capturing the underlying patterns of the hybrid-field channel without relying on explicit prior knowledge.
Specifically, we replace the proximal mapping step with a neural network module, termed proximal mapping module (PMM).
The PMM can be expressed as
\begin{equation}
\bar{\boldsymbol{g}}_{k} = \text{PMM}(\boldsymbol{z}_{k}).
\end{equation}
This module enables the CNN-CDL to overcome the limitations of predefined regularization functions and eliminates the need for explicitly performing the proximal mapping step.

The PMM utilizes a symmetric encoder-decoder architecture with three hierarchical levels, as illustrated in Fig.\ref{CNN_CDL}.
The module begins and ends with 3×3 convolutional layers. 
The encoder pathway consists of convolutional blocks (CB) and downsampling operations, where the downsampling is performed using convolutions with a kernel size of 2 and a stride of 2.  
The decoder pathway adopts a similar structure, employing CB and upsampling operation. 
These upsampling operations utilize point-wise convolutions and pixel shuffling. 
Skip connections connect corresponding layers in the encoder and decoder pathways by concatenating their channels, followed by point-wise convolutions.

The structure of CB is illustrated in Fig.\ref{3components}. 
Each CB comprises two sub-blocks, with each sub-block being preceded by layer normalization and having its own residual connection.
The first sub-block processes input features through two parallel branches: a static convolution path and a channel-spatial attention (CSA) path. 
The static path consists of a 1×1 pointwise convolution followed by a 3×3 depthwise convolution, while the CSA branch adaptively processes spatial and channel information. 
These branches combine via element-wise multiplication.
The second sub-block consists of two 1×1 convolutions with an intermediate PReLU activation.
\begin{figure*}[htbp]
    \centering
    \includegraphics[scale=0.9]{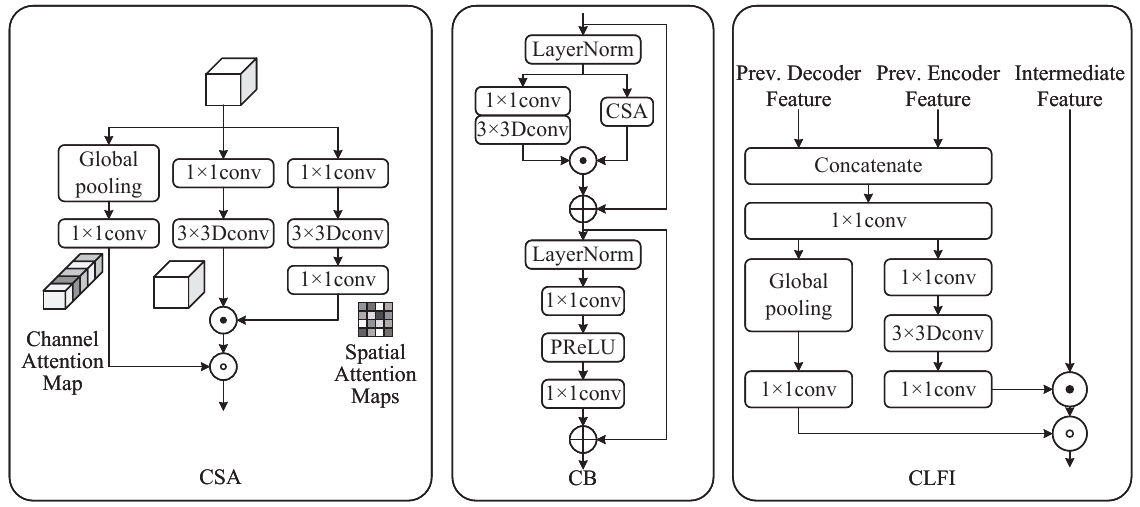}

    \caption{Detailed structures of the three key components: CSA, CB, and CLFI}
    \label{3components}
\end{figure*}

The CSA path plays a crucial role in enhancing feature representation within the CB.
While vanilla self-attention mechanism\cite{NIPS2017_3f5ee243} is effective, it requires the global computation of similarities between all features, leading to quadratic complexity with respect to the feature map size\cite{9912362}.  
This computational burden makes it impractical for high-dimensional channel estimation problems.
Inspired by advances in channel attention\cite{hu2018squeeze,chen2022simple}, the CSA offers an alternative approach that enables adaptive processing of both spatial and channel feature information.
The CSA operates by first generating two attention maps from the input feature. 
The channel attention map is derived through global average pooling followed by a 1×1 convolution. 
Simultaneously, the spatial attention map is computed using a sequence of one 1×1 convolution, one 3×3 depth-wise convolution, and another 1×1 convolution. 
In parallel with attention map generation, the input feature undergoes transformation through a 1×1 convolution followed by a depth-wise convolution. 
The transformed features are then refined by applying the spatial attention map through element-wise multiplication, followed by channel-wise multiplication using the channel attention map.
\subsection{Cross Layer Features Integration}
In the proposed CNN-CDL, the PMM in each layer utilizes skip connections to facilitate feature flow between the encoder and decoder within the same layer.
However, since each layer of the CNN-CDL corresponds to a single iteration of the PGD algorithm, they are relatively independent and primarily focused on the features of the current layer.
This lack of direct interaction with features from other layers might hinder the transfer of feature information across layers, thereby restricting overall performance.

To address this limitation, we introduce the cross layer feature integration (CLFI) block, as illustrated in Fig.\ref{3components}, to enhance the feature flow across layers.
At each scale, encoder and decoder features from the previous layer are transmitted to the current layer.
These features are first concatenated by channel and then processed by a 1×1 convolutional layer.
Similar to CSA, the resulting feature is used to compute two attention maps: a spatial attention map and a channel attention map.
These maps are then applied to refine the intermediate encoder feature. 
The spatial attention map first modulates the feature through element-wise multiplication, followed by channel-wise multiplication using the channel attention map.
By integrating features across layers, the PMM can leverage multi-scale and multi-layer information.

\subsection{Loss Function}
The proposed CNN-CDL aims to improve channel estimation performance through end-to-end training. 
To achieve this goal, we design a loss function that accounts for the estimation accuracy at each layer of the network.
The loss function is defined as
\begin{equation}
\mathcal{L}_{\text{CNN-CDL}} = \sum_{k=1}^{K} \left\Vert \mathbf{h} - \bar{\boldsymbol{g}}_{k}  \right\Vert_2^2.
\end{equation}
This loss function provides supervision at each layer, mitigating the vanishing gradient problem and encouraging progressive refinement of the channel estimate.
The CNN-CDL is trained by minimizing $\mathcal{L}$ with respect to all the learnable parameters of the CNN-CDL by employing batch stochastic gradient descent optimization.
This end-to-end training enables CNN-CDL to effectively capture the characteristics of hybrid-field channels and provide accurate channel estimates.

\section{Simulation Results}\label{sec:5}
\subsection{Simulation Setup}
To evaluate the performance of the proposed methods, we conduct simulations for XL-RIS-aided MIMO systems.
The system geometry is described using a Cartesian coordinate system, where the XL-RIS is positioned on the $yOz$-plane, with its center located at the origin $(0,0,0)$.
The region occupied by the XL-RIS is defined as
\begin{equation}
    \mathcal{S}_{\mathrm{R}}:=\left\{\left(x_r, y_r, z_r\right)|  x_r=0,| y_r|\leq \frac{L_y}{2},| z_r |, \leq \frac{L_z}{2}\right\},
\end{equation}
where $L_y$ and $L_z$ denote the dimensions of XL-RIS along the $y$ and $z$ axes, respectively.
Considering the UPA configuration for the XL-RIS with element spacing $d$, the dimensions are given by $L_y= M_1 d$ and $L_z=M_2 d$.
In our simulations, the XL-RIS consists of $M=M_1 \times M_2 = 64 \times 8$ elements.
Each element of the RIS phase shift matrix $\mathbf{\Theta}$ is randomly selected from the set$\left\{-{1}/{\sqrt{M}},+{1}/{\sqrt{M}}\right\}$.
The carrier frequency is set to 10 GHz, corresponding to a wavelength $\lambda = 0.03$ m.
Given this setup, the Rayleigh distance in this scenario is $62.4$ m.

The BS is positioned in the far-field region of the XL-RIS.
The BS is equipped with 32 antennas, and we consider $L_1 = 3$ propagation paths.  
For each path, the complex gain is randomly generated from a complex Gaussian distribution $\mathcal{CN}(0,1)$.
The departure angles $(\phi_i, \varphi_i)$ and the arrival angle $\psi_i$ are randomly generated from a uniform distribution over the interval $[-\pi/3, \pi/3]$.

The channel between the user and XL-RIS is modeled as a hybrid-field scenario, incorporating both far-field and near-field paths. 
For this hybrid-field scenario, we consider $L_2 = 6$ propagation paths, with an equal split between far-field and near-field paths.
For the far-field channel, the channel parameters are set similarly to those of the BS-RIS channel, including the distribution of gains and angles.
For the near-field channel, the scatterers are randomly distributed within a cuboid region defined by
\begin{align}
\mathcal{S}_{\mathrm{NF}} := \{&(x_s, y_s, z_s) \,|\, 3\text{ m} \leq x_s \leq 5\text{ m}, -15\text{ m} \leq y_s \leq 10\text{ m}, \nonumber \\
&-10.5\text{ m} \leq z_s \leq 0\text{ m}\}.
\end{align}
This region ensures all scatterers remain within the Rayleigh distance, preserving the near-field characteristics.
\subsection{Performance Evaluation}
We compare our proposed methods with the following baseline methods:
\begin{itemize}
    \item OMP: orthogonal matching pursuit method using a pre-defined dictionary from \cite{9810144}.
     \item ISTA-Net+: an unrolled network architecture based on the ISTA \cite{zhang2018ista}.
     \item OLS: oracle least squares method which assumes perfect knowledge of channel parameters and serves as a performance upper bound.
\end{itemize}

To assess computational complexity, we employ the fvcore library\footnote{The package is available at: https://github.com/facebookresearch/fvcore}  to count the number of floating-point operations (FLOPs) required by each method.
Table \ref{tab:method_comparison} summarizes the FLOPs and dictionary sizes for each method.
\begin{table}
\caption{Comparison of Dictionary Sizes and Computational Complexity}
\label{tab:method_comparison}
\centering
\begin{tabular}{|c|c|c|}
\hline
Method & Dictionary Size & FLOPs ($\times 10^9$) \\
\hline
OMP & $101475 \times 512$ & 0.51 \\
ISTA-Net+ & - & 8.47 \\
CISTA-Net & $608 \times 1$ & 0.54 \\
CISTA-Net+ & $608 \times 13 \times 2$ & 0.54 \\
CNN-CDL & $ 1,186 \times 5\times2$ & 25.42 \\
\hline
\end{tabular}
\end{table}
The dictionary structures of our proposed methods are as follows: CISTA-Net and CISTA-Net+ employ a convolutional dictionary of size $(2, 32, 3\times 3)$, resulting in 608 parameters. For CISTA-Net+, due to the use of decoupled layer-specific convolutional dictionaries, the total becomes $608 \times 13 \times 2$, where 13 is the number of network layers and 2 comes from the decoupling of the dictionary.
For CNN-CDL, the CNN block structure $\mathcal{F}_{k}^{D}$ and $\mathcal{F}_{k}^{A}$ each consists of convolutional layers of sizes $(2, 32, 3\times 3)$ and $(32, 2, 3\times 3)$, resulting in 1,186 parameters each. Thus, the overall count is $1,186 \times 5 \times 2$, where 5 is the number of network layers and 2 again represents the decoupled CNN blocks.
The performance is evaluated in terms of normalized mean squared error (NMSE), where $\hat{\mathbf{h}}$ is the estimated channel
\begin{equation}
\mathrm{NMSE}=\mathbb{E}\left\{\|\hat{\mathbf{h}}-\mathbf{h}\|_2^2 /\|\mathbf{h}\|_2^2\right\}.
\end{equation}

We first evaluate the performance of our proposed methods in terms of NMSE versus SNR, with the number of pilots fixed at 256. Fig. \ref{fig:nmse_vs_snr} illustrates the comparison results.
CISTA-Net consistently outperforms OMP across all SNR levels. 
This improvement demonstrates the advantage of incorporating CDL for hybrid-field channel estimation. 
Unlike traditional compressed sensing methods with predefined dictionary matrices, CISTA-Net more effectively captures channel characteristics.
Building upon CISTA-Net, CISTA-Net+ shows further enhancement in performance. 
This improvement stems from three key modifications: the incorporation of learnable parameters, the separation of measurement matrices and dictionaries, and the adoption of a supervised learning approach. These changes enable CISTA-Net+ to better adapt to the complex structures inherent in hybrid-field channels, resulting in consistent performance gains across the entire SNR range when compared to ISTA-Net+.
Among all methods evaluated, CNN-CDL achieves the best performance, with NMSE values very close to those of OLS. This superior performance can be attributed to two key innovations: the integration of CNNs into the CDL and the introduction of a data-driven proximal mapping. The CNN structure allows for more effective capture of complex channel features, while the data-driven proximal mapping adapts to the intricate structures of hybrid-field channels without relying on explicit prior knowledge. 
\begin{figure}[htbp]
\centering
\includegraphics[width=\columnwidth]{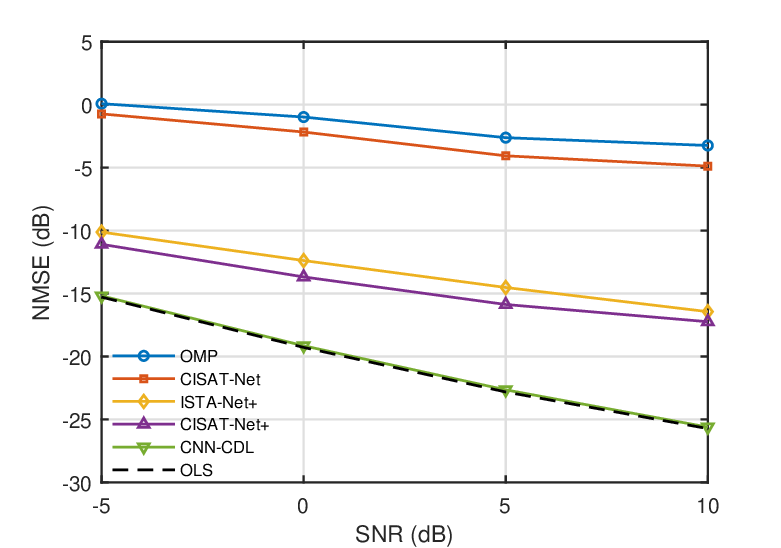}
\caption{NMSE performance comparison versus SNR with 256 pilots}
\label{fig:nmse_vs_snr}
\end{figure}

Similar trends are observed in Fig. \ref{fig:nmse_vs_snr_128}, where we modified the antenna architecture. Specifically, the RIS configuration is changed from 512 $(64\times8)$ elements to 256 $(64\times4)$ elements. 
Additionally, the number of BS antennas is reduced from 32 to 16, and the number of pilots is decreased from 256 to 128. 
It is worth noting that with these changes, the Rayleigh distance only slightly changed from 62.5m to 61.6m. 
This minimal change in distance allows the same scatterer distribution for the near-field components to be maintained.
At an SNR of 0 dB, CNN-CDL achieves an NMSE of -12.08 dB, which is close to the OLS upper bound, with only a 0.11 dB difference. 
CNN-CDL outperforms CISTA-NET+ by 2.75 dB, with CISTA-NET+ achieving an NMSE of -9.33 dB. 
In turn, CISTA-NET+ shows improvements over ISTA-NET+, surpassing ISTA-NET by 0.95 dB.
CISTA-NET itself demonstrates a 1.23 dB advantage over OMP, which achieves an NMSE of -0.34 dB. 
These results highlight the progressive improvements achieved by each method in the proposed series.
\begin{figure}[htbp]
\centering
\includegraphics[width=\columnwidth]{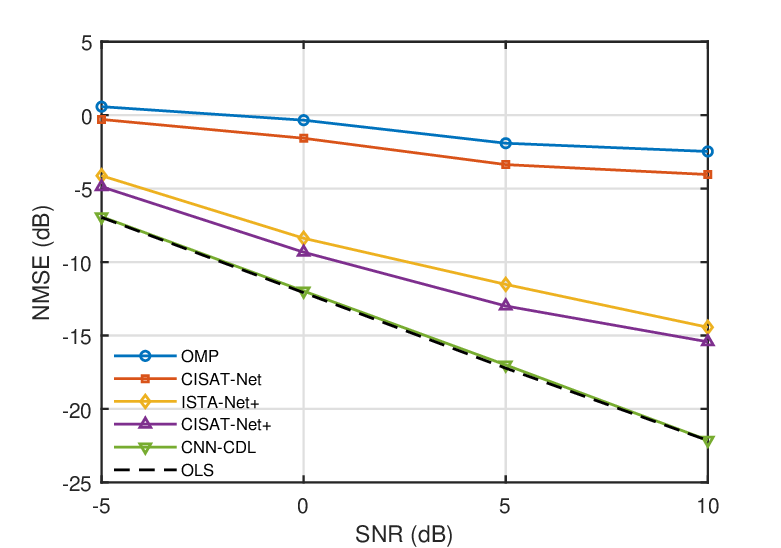}
\caption{NMSE performance comparison versus SNR with 128 pilots}
\label{fig:nmse_vs_snr_128}
\end{figure}

Fig. \ref{fig:nmse_vs_pilots} illustrates the performance comparison in terms of NMSE versus the number of pilots at an SNR of 0 dB, with the RIS configuration fixed at  $64\times8$ elements. 
The number of pilots varies among  64, 128, 192, and 256.
The results demonstrate that as the number of pilots decreases, the available observation information becomes more limited, leading to performance degradation across all methods.
As the number of pilots increases, the performance gap between CISTA-Net+ and ISTA-Net+ also increases. 
This trend indicates that CISTA-Net+ is more effective at utilizing the increased observation information to improve channel estimation. 
Consequently, this highlights the superior capability of the proposed CDL in learning and capturing complex channel characteristics.
Among the evaluated methods, CNN-CDL consistently outperforms others across all pilot numbers, with its performance closely approaching that of OLS.
This capability is attributed to the introduction of CNN and data-driven proximal mapping modules, effectively capturing complex hybrid-field channel characteristics with limited pilots.

\begin{figure}[htbp]
\centering
\includegraphics[width=\columnwidth]{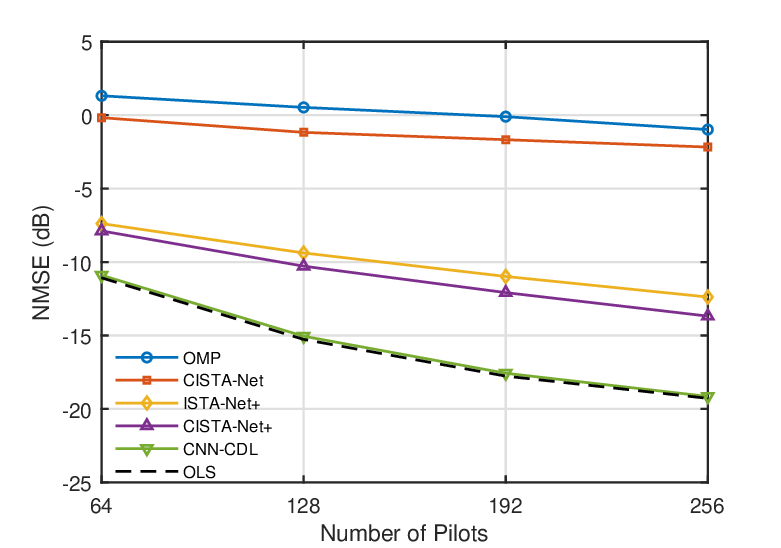 }
\caption{NMSE performance comparison versus number of pilots at 0 dB SNR}
\label{fig:nmse_vs_pilots}
\end{figure}

To evaluate the robustness of the proposed methods under different channel conditions, we investigate their performance across multiple multipath configurations. 
While maintaining a total of $L_2=6$ propagation paths, we adjust the ratio of far-field to near-field paths. The multipath configurations include [0,6] (purely near-field), [1,5], [3,3], [5,1], and [6,0] (purely far-field).
Fig. \ref{fig:nmse_vs_path} illustrates the NMSE performance of different methods under these multipath configurations. 
OMP shows a performance improvement as the number of near-field paths increases, but deteriorates with more far-field paths. This behavior can be attributed to OMP's reliance on a pre-defined dictionary that is primarily tailored for near-field scenarios, thus performing better when the channel characteristics align with these assumptions.
For our proposed CISTA-Net, CISTA-Net+, and CNN-CDL, we consider two training scenarios: (1) retraining for each path configuration, and (2) training only on the [3,3] configuration and testing on others. In the first scenario, all three methods demonstrate relatively stable performance across different path configurations, showcasing their adaptability to varying channel conditions.
However, the second scenario reveals variations among the methods. CISTA-Net exhibits optimal performance at [3,3] but degrades significantly at both extremes. 
CISTA-Net+ shows improved robustness compared to CISTA-Net, while CNN-CDL demonstrates the best overall performance. 
These results highlight the superior adaptability of CNN-CDL to varying multipath scenarios, even when trained on a single configuration. This robustness can be attributed to its advanced neural network architecture, which effectively captures both near-field and far-field channel characteristics.
\begin{figure}[htbp]
\centering
\includegraphics[width=\columnwidth]{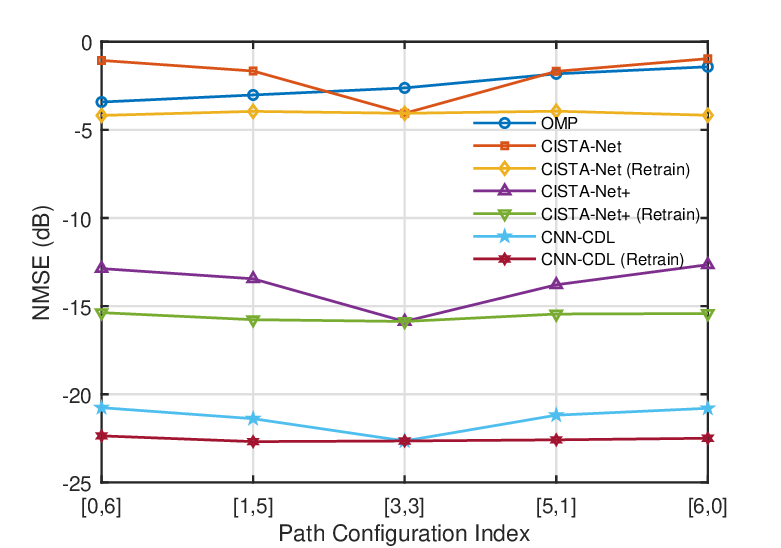}
\caption{NMSE performance under different multipath configurations at 5 dB SNR}
\label{fig:nmse_vs_path}
\end{figure}

To assess the practical impact of channel estimation accuracy on system performance, we compare the spectral efficiency (SE) based on perfect CSI and estimated channels. 
The SE is defined as follow
\begin{equation}
R= \log _2\left(1+\frac{W}{\sigma_n^2}\left|\mathbf{f}^{H}\mathbf{H} \boldsymbol{\theta} \right|^2\right),
\end{equation}
where $W$ is the transmit power and $\mathbf{f}$ is the digital beamforming vector.
We consider three cases: perfect CSI, CNN-CDL channel estimation, and CISTA-Net+ channel estimation. 
For each case, we employ the gradient projection method \cite{10477515} to solve the SE optimization problem.
Fig. \ref{fig:SE_vs_power} illustrates the SE versus transmit power for these three cases. 
As expected, the perfect CSI achieves the highest SE across all power levels, serving as an upper bound for performance. 
CNN-CDL demonstrates superior performance compared to CISTA-Net+, with its SE curve lying closer to the perfect CSI case.
\begin{figure}[htbp]
\centering
\includegraphics[width=\columnwidth]{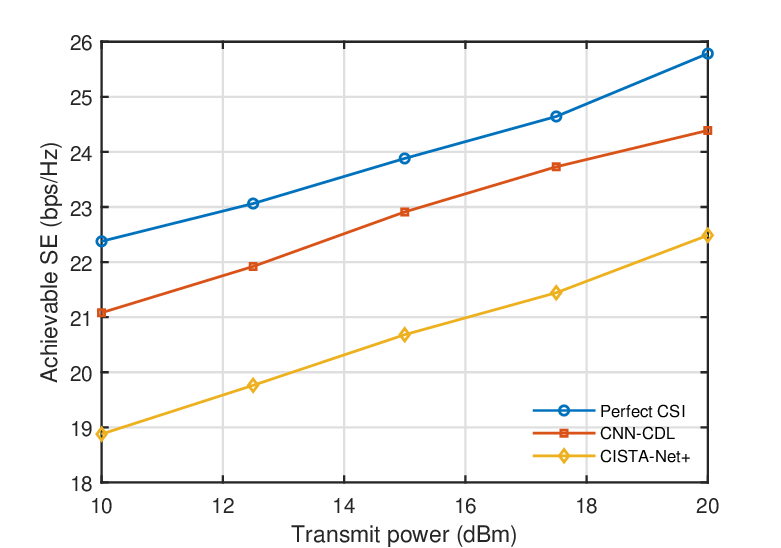}
\caption{Achievable Spectral Efficiency versus transmit power for perfect CSI and estimated channels}
\label{fig:SE_vs_power}
\end{figure}

Fig. \ref{fig:convergence_speed} illustrates the NMSE performance versus the number of unrolled layers for each method. 
The results demonstrate differences in convergence rates.
CNN-CDL exhibits the fastest convergence, achieving an NMSE of -25.61 dB at the 5-th layer. 
CISTA-Net+ shows improved performance over the CISTA-Net, reaching an NMSE of -17.24 dB at the 13-th layer, while CISTA-Net attains an NMSE of -4.88 dB at the 17-th layer.
\begin{figure}[htbp]
\centering
\includegraphics[width=\columnwidth]{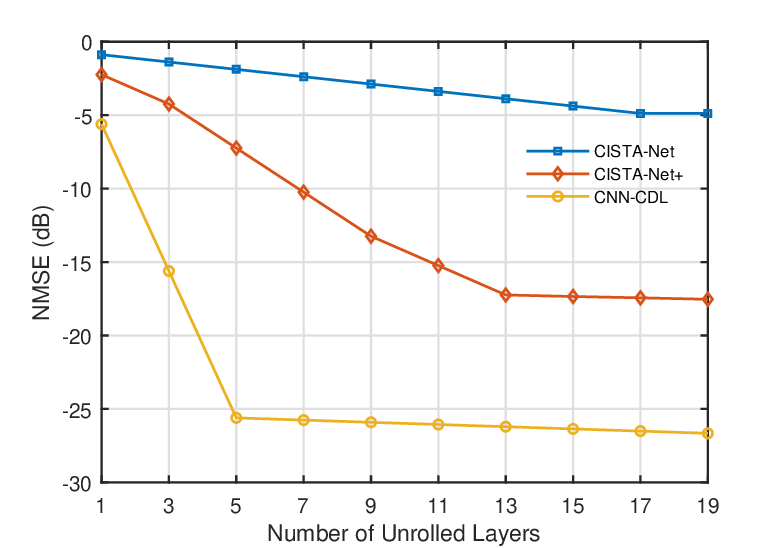}
\caption{NMSE performance versus number of unrolled layers for different methods}
\label{fig:convergence_speed}
\end{figure}
\section{Conclusion}\label{sec:6}
In this work, we addressed the challenge of channel estimation in XL-RIS-aided MIMO systems, specifically focusing on hybrid-field scenarios. We approached this problem through the lens of dictionary learning, which allows for optimization of the dictionary and estimated channel.
To handle the high-dimensional nature of XL-RIS channels, we specifically adopt the CDL formulation.
We cast the CDL formulation as a bilevel optimization problem and adopt a gradient-based approach to solve it. To address the challenges in computing gradients, we introduced an unrolled optimization method based on the PGD algorithm. This algorithm simplifies to the ISTA algorithm. Building upon the ISTA, we developed CISTA-Net and its enhanced version CISTA-Net+, which transform ISTA into learnable layers for efficient end-to-end CDL optimization.
To further enhance performance, we proposed the CNN-CDL approach. This method incorporates deep learning components into the PGD algorithm, including CNN blocks and a learnable proximal mapping module. We also integrated cross-layer features to leverage multi-scale and multi-layer information.
Simulation results demonstrate the effectiveness of the proposed methods for channel estimation in hybrid-field XL-RIS-aided massive MIMO systems.


%





\ifCLASSOPTIONcaptionsoff
  \newpage
\fi



\bibliographystyle{IEEEtran}
\bibliography{IEEEabrv,IEEEexample}

\begin{thebibliography}{10}
\providecommand{\url}[1]{#1}
\csname url@samestyle\endcsname
\providecommand{\newblock}{\relax}
\providecommand{\bibinfo}[2]{#2}
\providecommand{\BIBentrySTDinterwordspacing}{\spaceskip=0pt\relax}
\providecommand{\BIBentryALTinterwordstretchfactor}{4}
\providecommand{\BIBentryALTinterwordspacing}{\spaceskip=\fontdimen2\font plus
\BIBentryALTinterwordstretchfactor\fontdimen3\font minus \fontdimen4\font\relax}
\providecommand{\BIBforeignlanguage}[2]{{%
\expandafter\ifx\csname l@#1\endcsname\relax
\typeout{** WARNING: IEEEtran.bst: No hyphenation pattern has been}%
\typeout{** loaded for the language `#1'. Using the pattern for}%
\typeout{** the default language instead.}%
\else
\language=\csname l@#1\endcsname
\fi
#2}}
\providecommand{\BIBdecl}{\relax}
\BIBdecl

\bibitem{9424177}
Y.~Liu, X.~Liu, X.~Mu, T.~Hou, J.~Xu, M.~Di~Renzo, and N.~Al-Dhahir, ``Reconfigurable intelligent surfaces: Principles and opportunities,'' \emph{IEEE Communications Surveys {\&} Tutorials}, vol.~23, no.~3, pp. 1546--1577, 2021.

\bibitem{9722893}
B.~Zheng, C.~You, W.~Mei, and R.~Zhang, ``A survey on channel estimation and practical passive beamforming design for intelligent reflecting surface aided wireless communications,'' \emph{IEEE Communications Surveys {\&} Tutorials}, vol.~24, no.~2, pp. 1035--1071, 2022.

\bibitem{9771077}
A.~L. Swindlehurst, G.~Zhou, R.~Liu, C.~Pan, and M.~Li, ``Channel estimation with reconfigurable intelligent surfaces—{A} general framework,'' \emph{Proceedings of the IEEE}, vol. 110, no.~9, pp. 1312--1338, 2022.

\bibitem{9732214}
G.~Zhou, C.~Pan, H.~Ren, P.~Popovski, and A.~L. Swindlehurst, ``Channel estimation for {RIS}-aided multiuser millimeter-wave systems,'' \emph{IEEE Transactions on Signal Processing}, vol.~70, pp. 1478--1492, 2022.

\bibitem{9328485}
X.~Wei, D.~Shen, and L.~Dai, ``Channel estimation for {RIS} assisted wireless communications—part {II}: An improved solution based on double-structured sparsity,'' \emph{IEEE Communications Letters}, vol.~25, no.~5, pp. 1403--1407, 2021.

\bibitem{10053657}
J.~Chen, Y.-C. Liang, H.~V. Cheng, and W.~Yu, ``Channel estimation for reconfigurable intelligent surface aided multi-user {mmWave} {MIMO} systems,'' \emph{IEEE Transactions on Wireless Communications}, vol.~22, no.~10, pp. 6853--6869, 2023.

\bibitem{9881980}
Y.~Guo, P.~Sun, Z.~Yuan, C.~Huang, Q.~Guo, Z.~Wang, and C.~Yuen, ``Efficient channel estimation for {RIS}-aided {MIMO} communications with unitary approximate message passing,'' \emph{IEEE Transactions on Wireless Communications}, vol.~22, no.~2, pp. 1403--1416, 2023.

\bibitem{9103231}
P.~Wang, J.~Fang, H.~Duan, and H.~Li, ``Compressed channel estimation for intelligent reflecting surface-assisted millimeter wave systems,'' \emph{IEEE Signal Processing Letters}, vol.~27, pp. 905--909, 2020.

\bibitem{9475488}
H.~Liu, J.~Zhang, Q.~Wu, H.~Xiao, and B.~Ai, ``{ADMM} based channel estimation for {RISs} aided millimeter wave communications,'' \emph{IEEE Communications Letters}, vol.~25, no.~9, pp. 2894--2898, 2021.

\bibitem{9521836}
Z.~Chen, J.~Tang, X.~Y. Zhang, D.~K.~C. So, S.~Jin, and K.-K. Wong, ``Hybrid evolutionary-based sparse channel estimation for {IRS}-assisted {mmWave} {MIMO} systems,'' \emph{IEEE Transactions on Wireless Communications}, vol.~21, no.~3, pp. 1586--1601, 2022.

\bibitem{spawc}
P.~Zheng, X.~Lyu, Y.~Wang, and Y.~Gong, ``Dictionary learning based near-field channel estimation for wideband {XL-MIMO} systems,'' in \emph{2024 IEEE 25th International Workshop on Signal Processing Advances in Wireless Communications (SPAWC)}, 2024, pp. 1--5.

\bibitem{9810144}
X.~Wei, L.~Dai, Y.~Zhao, G.~Yu, and X.~Duan, ``Codebook design and beam training for extremely large-scale {RIS}: Far-field or near-field?'' \emph{China Communications}, vol.~19, no.~6, pp. 193--204, 2022.

\bibitem{10496996}
H.~Lu, Y.~Zeng, C.~You, Y.~Han, J.~Zhang, Z.~Wang, Z.~Dong, S.~Jin, C.-X. Wang, T.~Jiang, X.~You, and R.~Zhang, ``A tutorial on near-field {XL-MIMO} communications towards {6G},'' \emph{IEEE Communications Surveys {\&} Tutorials}, pp. 1--1, 2024.

\bibitem{10178011}
P.~Zheng, X.~Lyu, and Y.~Gong, ``Trainable proximal gradient descent-based channel estimation for {mmWave} massive {MIMO} systems,'' \emph{IEEE Wireless Communications Letters}, vol.~12, no.~10, pp. 1781--1785, 2023.

\bibitem{10379539}
Z.~Wang, J.~Zhang, H.~Du, D.~Niyato, S.~Cui, B.~Ai, M.~Debbah, K.~B. Letaief, and H.~V. Poor, ``A tutorial on extremely large-scale {MIMO} for {6G}: Fundamentals, signal processing, and applications,'' \emph{IEEE Communications Surveys {\&} Tutorials}, pp. 1--1, 2024.

\bibitem{9693928}
M.~Cui and L.~Dai, ``Channel estimation for extremely large-scale {MIMO}: Far-field or near-field?'' \emph{IEEE Transactions on Communications}, vol.~70, no.~4, pp. 2663--2677, 2022.

\bibitem{10149498}
J.~Wu, S.~Kim, and B.~Shim, ``Parametric sparse channel estimation for {RIS}-assisted terahertz systems,'' \emph{IEEE Transactions on Communications}, vol.~71, no.~9, pp. 5503--5518, 2023.

\bibitem{10081022}
S.~Yang, W.~Lyu, Z.~Hu, Z.~Zhang, and C.~Yuen, ``Channel estimation for near-field {XL-RIS}-aided {mmWave} hybrid beamforming architectures,'' \emph{IEEE Transactions on Vehicular Technology}, vol.~72, no.~8, pp. 11\,029--11\,034, 2023.

\bibitem{10464973}
Z.~Tang, Y.~Chen, Y.~Wang, T.~Mao, Q.~Wu, M.~Di~Renzo, and L.~Hanzo, ``Near-field sparse channel estimation for extremely large-scale {RIS}-aided wireless communications,'' in \emph{2023 IEEE Globecom Workshops (GC Wkshps)}, 2023, pp. 1373--1379.

\bibitem{10153711}
X.~Yu, W.~Shen, R.~Zhang, C.~Xing, and T.~Q.~S. Quek, ``Channel estimation for {XL-RIS}-aided millimeter-wave systems,'' \emph{IEEE Transactions on Communications}, vol.~71, no.~9, pp. 5519--5533, 2023.

\bibitem{10077727}
J.~Xiao, J.~Wang, Z.~Chen, and G.~Huang, ``{U-MLP}-based hybrid-field channel estimation for {XL-RIS} assisted millimeter-wave {MIMO} systems,'' \emph{IEEE Wireless Communications Letters}, vol.~12, no.~6, pp. 1042--1046, 2023.

\bibitem{9598863}
X.~Wei and L.~Dai, ``Channel estimation for extremely large-scale massive {MIMO}: Far-field, near-field, or hybrid-field?'' \emph{IEEE Communications Letters}, vol.~26, no.~1, pp. 177--181, 2022.

\bibitem{9940281}
Z.~Hu, C.~Chen, Y.~Jin, L.~Zhou, and Q.~Wei, ``Hybrid-field channel estimation for extremely large-scale massive {MIMO} system,'' \emph{IEEE Communications Letters}, vol.~27, no.~1, pp. 303--307, 2023.

\bibitem{10013010}
H.~Nayir, E.~Karakoca, A.~Görçin, and K.~Qaraqe, ``Hybrid-field channel estimation for massive {MIMO} systems based on {OMP} cascaded convolutional autoencoder,'' in \emph{2022 IEEE 96th Vehicular Technology Conference (VTC2022-Fall)}, 2022, pp. 1--6.

\bibitem{10546479}
H.~Lei, J.~Zhang, Z.~Wang, B.~Ai, and D.~W. Kwan~Ng, ``Hybrid-field channel estimation for {XL-MIMO} systems with stochastic gradient pursuit algorithm,'' \emph{IEEE Transactions on Signal Processing}, vol.~72, pp. 2998--3012, 2024.

\bibitem{10143629}
W.~Yu, Y.~Shen, H.~He, X.~Yu, S.~Song, J.~Zhang, and K.~B. Letaief, ``An adaptive and robust deep learning framework for {THz} ultra-massive {MIMO} channel estimation,'' \emph{IEEE Journal of Selected Topics in Signal Processing}, vol.~17, no.~4, pp. 761--776, 2023.

\bibitem{8383706}
Y.~Ding and B.~D. Rao, ``Dictionary learning-based sparse channel representation and estimation for {FDD} massive {MIMO} systems,'' \emph{IEEE Transactions on Wireless Communications}, vol.~17, no.~8, pp. 5437--5451, 2018.

\bibitem{9186336}
H.~Xie and N.~González-Prelcic, ``Dictionary learning for channel estimation in hybrid frequency-selective {mmWave} {MIMO} systems,'' \emph{IEEE Transactions on Wireless Communications}, vol.~19, no.~11, pp. 7407--7422, 2020.

\bibitem{10054604}
H.~Xie, J.~Palacios, and N.~González-Prelcic, ``Hybrid {mmWave} {MIMO} systems under hardware impairments and beam squint: Channel model and dictionary learning-aided configuration,'' \emph{IEEE Transactions on Wireless Communications}, vol.~22, no.~10, pp. 6898--6913, 2023.

\bibitem{10197339}
Y.~Zhao, Y.~Teng, A.~Liu, X.~Wang, and V.~K.~N. Lau, ``Joint {UL/DL} dictionary learning and channel estimation via two-timescale optimization in massive {MIMO} systems,'' \emph{IEEE Transactions on Wireless Communications}, vol.~23, no.~3, pp. 2369--2382, 2024.

\bibitem{9919846}
Z.~Peng, G.~Zhou, C.~Pan, H.~Ren, A.~L. Swindlehurst, P.~Popovski, and G.~Wu, ``Channel estimation for {RIS}-aided multi-user {mmWave} systems with uniform planar arrays,'' \emph{IEEE Transactions on Communications}, vol.~70, no.~12, pp. 8105--8122, 2022.

\bibitem{6717211}
O.~E. Ayach, S.~Rajagopal, S.~Abu-Surra, Z.~Pi, and R.~W. Heath, ``Spatially sparse precoding in millimeter wave {MIMO} systems,'' \emph{IEEE Transactions on Wireless Communications}, vol.~13, no.~3, pp. 1499--1513, 2014.

\bibitem{10123941}
Z.~Wu and L.~Dai, ``Multiple access for near-field communications: {SDMA} or {LDMA}?'' \emph{IEEE Journal on Selected Areas in Communications}, vol.~41, no.~6, pp. 1918--1935, 2023.

\bibitem{5714407}
I.~Tošić and P.~Frossard, ``Dictionary learning,'' \emph{IEEE Signal Processing Magazine}, vol.~28, no.~2, pp. 27--38, 2011.

\bibitem{1710377}
M.~Aharon, M.~Elad, and A.~Bruckstein, ``{K-SVD}: An algorithm for designing overcomplete dictionaries for sparse representation,'' \emph{IEEE Transactions on Signal Processing}, vol.~54, no.~11, pp. 4311--4322, 2006.

\bibitem{5466111}
J.~Yang, J.~Wright, T.~S. Huang, and Y.~Ma, ``Image super-resolution via sparse representation,'' \emph{IEEE Transactions on Image Processing}, vol.~19, no.~11, pp. 2861--2873, 2010.

\bibitem{8364626}
C.~Garcia-Cardona and B.~Wohlberg, ``Convolutional dictionary learning: A comparative review and new algorithms,'' \emph{IEEE Transactions on Computational Imaging}, vol.~4, no.~3, pp. 366--381, 2018.

\bibitem{plaut2019greedy}
E.~Plaut and R.~Giryes, ``A greedy approach to {$\ell_{0,\infty}$} -based convolutional sparse coding,'' \emph{SIAM {Journal} on {Imaging} {Sciences}}, vol.~12, no.~1, pp. 186--210, 2019.

\bibitem{10502023}
Y.~Zhang, P.~Khanduri, I.~Tsaknakis, Y.~Yao, M.~Hong, and S.~Liu, ``An introduction to {Bilevel} optimization: Foundations and applications in signal processing and machine learning,'' \emph{IEEE Signal Processing Magazine}, vol.~41, no.~1, pp. 38--59, 2024.

\bibitem{parikh2014proximal}
N.~Parikh and S.~Boyd, ``Proximal algorithms,'' \emph{Foundations and trends{\textregistered} in Optimization}, vol.~1, no.~3, pp. 127--239, 2014.

\bibitem{beck2009fast}
A.~Beck and M.~Teboulle, ``A fast iterative shrinkage-thresholding algorithm for linear inverse problems,'' \emph{SIAM journal on imaging sciences}, vol.~2, no.~1, pp. 183--202, 2009.

\bibitem{9363511}
V.~Monga, Y.~Li, and Y.~C. Eldar, ``Algorithm unrolling: Interpretable, efficient deep learning for signal and image processing,'' \emph{IEEE Signal Processing Magazine}, vol.~38, no.~2, pp. 18--44, 2021.

\bibitem{NIPS2017_3f5ee243}
A.~Vaswani, N.~Shazeer, N.~Parmar, J.~Uszkoreit, L.~Jones, A.~N. Gomez, L.~u. Kaiser, and I.~Polosukhin, ``Attention is all you need,'' in \emph{Advances in Neural Information Processing Systems}, vol.~30, 2017.

\bibitem{9912362}
M.-H. Guo, Z.-N. Liu, T.-J. Mu, and S.-M. Hu, ``Beyond self-attention: External attention using two linear layers for visual tasks,'' \emph{IEEE Transactions on Pattern Analysis and Machine Intelligence}, vol.~45, no.~5, pp. 5436--5447, 2023.

\bibitem{hu2018squeeze}
J.~Hu, L.~Shen, and G.~Sun, ``Squeeze-and-excitation networks,'' in \emph{Proceedings of the IEEE conference on Computer Vision and Pattern Recognition}, 2018, pp. 7132--7141.

\bibitem{chen2022simple}
L.~Chen, X.~Chu, X.~Zhang, and J.~Sun, ``Simple baselines for image restoration,'' in \emph{European {C}onference on {C}omputer {V}ision}, 2022, pp. 17--33.

\bibitem{zhang2018ista}
J.~Zhang and B.~Ghanem, ``{ISTA-Net}: Interpretable optimization-inspired deep network for image compressive sensing,'' in \emph{Proceedings of the IEEE conference on Computer Vision and Pattern Recognition}, 2018, pp. 1828--1837.

\bibitem{10477515}
X.~He, L.~Huang, J.~Wang, and Y.~Gong, ``Learn to optimize {RIS} aided hybrid beamforming with out-of-distribution generalization,'' \emph{IEEE Transactions on Vehicular Technology}, vol.~73, no.~7, pp. 10\,783--10\,787, 2024.

\end{thebibliography}
\end{document}